# ARQ Secrecy: From Theory to Practice

Yara Omar, Moustafa Youssef
Wireless Intelligent Networks Center (WINC)
Nile University, Cairo, Egypt
{yara.abdallah, mayoussef}@nileu.edu.eg

Hesham El Gamal
Department of Electrical and Computer Engineering
Ohio State University, Columbus, USA
helgamal@ece.osu.edu

*Abstract*—Inspired by our earlier work on Automatic Repeat reQuest (ARQ) secrecy, we propose a simple, yet efficient, security overlay protocol to existing 802.11 networks. Our work targets networks secured by the Wired Equivalent Privacy (WEP) protocol because of its widespread use and vulnerability to a multitude of security threats. By exploiting the existing ARQ protocol in the 802.11 standard, our proposed opportunistic secrecy scheme is shown to defend against all known passive WEP attacks. Moreover, our implementation on the madwifi-ng driver is used to establish the achievability of a vanishing secrecy outage probability in several realistic scenarios.

## I. INTRODUCTION

Wi-Fi networks are gaining popularity, in a wide variety of applications, at an exponential rate. Despite its vulnerability to several known security attacks and the availability of more robust security protocols, i.e., WPA and WPA2, the Wired Equivalence Privacy (WEP) protocol is still widely used for securing Wi-Fi networks. As of October 2008, RSA, The Security Division of EMC, reported that 48%, 38% and 24% of NYC, London and Paris Wi-Fi networks still employ the WEP protocol, respectively [18]. The reason is, arguably, the significant advantage that the WEP protocol enjoys, in terms of user friendliness, over competing approaches. This observation motivates our work. More specifically, we propose a novel variant of the WEP protocol that is capable of overcoming all passive WEP attacks, in a strong probabilistic sense which will be made rigorous in the sequel, while preserving the same user interface as the traditional WEP protocol. The proposed ARQ-WEP protocol is inspired by information theoretic principles and exploits the existing ARQ feedback adopted in the 802.11 standard.

Recent years have witnessed a resurgence in information theoretic security research. This line of work has mostly targeted wireless networking applications and has been largely inspired by Wyner's pioneering work on the wiretap channel [1], [2]. In this model, two users wish to securely communicate over a noisy channel in the presence of an eavesdropper; also impaired by a noisy channel. Quite remarkably, a non-zero secrecy capacity was shown to be achievable in [1], **without any limiting assumptions on the eavesdropper computational power or the availability of a private key**, if the wiretapper channel is a degraded version of the legitimate one. Csiszár and Körner later extended this result to the non-degraded channel scenario [3]. More recently, in [4], [5], the effect of slow fading on the secrecy capacity was studied, where it was shown that the secrecy capacity could be enhanced by distributing the secure message across different fading realizations. Of particular relevance to this paper is the approach proposed in [6] for exchanging secure messages using Hybrid ARQ protocols. This technique, however, suffers from a secrecy outage phenomenon resulting from the allocation of each secure message to only one ARQ epoch. Our work [7] develops a unified framework for sharing keys using ARQ which overcomes the secrecy outage phenomenon and lends itself to low complexity implementations. In fact, under the assumption of an error-free and public ARQ feedback, our results established the achievability of non-zero key rates, with a vanishing probability of secrecy outage, even when the eavesdropper is enjoying a higher average signal-to-noise ratio (SNR) than the legitimate receiver [7]. The *seemingly* counter-intuitive result is another manifestation of the opportunistic secrecy principle formulated in [4].

Building on this information theoretic foundation, this paper proposes a practical security protocol for Wi-Fi networks. Instead of aiming at a complete overhaul of the security mechanism adopted in the 802.11 standard, our **overlay** protocol only requires minor modifications in the current implementation of the popular WEP protocol. The resulting ARQ-WEP protocol, however, offers significant advantages over the competing approaches. On one hand, it inherits the simplicity of implementation and user friendliness from the WEP protocol (as opposed to the WPA and WPA2 protocols). The ARQ-WEP protocol, on the other hand, achieves significant security gains relative to the traditional implementation of the WEP protocol. More specifically, our analytical and experimental results reveal the ability of the ARQ-WEP protocol to defend against all passive WEP attacks at the expense of a slight degradation in throughput over the original WEP protocol.

The rest of this paper is organized as follows. Section II discusses the main ingredients of the WEP protocol and identifies its vulnerabilities. In Section III, the basic idea behind ARQ security is introduced. The ARQ-WEP protocol, is developed in Section IV. Section V demonstrates the significant security advantage offered by the ARQ-WEP protocol via our experimental results. Finally, we offer some concluding remarks in Section VI.

## II. THE WEP PROTOCOL

We now review the basic setup of the WEP security protocol. Our overarching goal is to unveil the main weakness of the protocol, and hence, motivate our information theoretic solution based on ARQ secrecy. Here, a sender node (Alice) sends a number of encrypted data frames, $n_d$, to a receiver node (Bob), in a Wi-Fi network secured by WEP. Both nodes follow the ARQ mechanism adopted in the IEEE 802.11 standard.

Alice and Bob are assumed to share a single 104-bit **private** key, $K_s$, which is used to encrypt/decrypt all $n_d$ data frames sent by Alice. For the $i^{th}$ data frame, containing a message $M(i)$, a CRC32 checksum (used as an Integrity Check Value), i.e., $ICV(M(i))$, is computed and appended to the message forming the plaintext, $P(i)$. The RC4 algorithm is then seeded with the concatenation of a pseudo-random 24-bit Initialization Vector (IV), denoted by $V(i)$, and the private key to generate the keystream, $RC4(V(i), K_s)$. The ciphertext, $C(i)$, is obtained by XORing the plaintext with the generated keystream. Finally, Alice sends $C(i)$ along with the IV. More formally,

$$C(i) = P(i) \bigoplus RC4(V(i), K_s), \quad (1)$$
$$A \rightarrow B : V(i), C(i). \quad (2)$$

After recovering the IV, which was sent as plaintext in the MAC header, Bob generates the RC4 keystream, $RC4(V(i), K_s)$, and XORes it with the received ciphertext to obtain the plaintext, $P(i)$. The final step is to compute a checksum, $ICV'$, from the received message, $M(i)$, and compare it with the received $ICV'(i)$. If they match, successful decryption is declared and the frame is passed to higher layers; otherwise an error is declared and the frame is dropped.

A brief overview of the history of WEP attacks is now in order. WEP design failures were first reported by Borisov, Goldberg, and Wagner in 2001 [11]. They showed that the ICV cannot be used against malicious attacks. Additionally, they observed that old IVs could be reused, allowing for a straightforward path for packet injection. During the same year, Fluhrer, Mantin and Shamir presented the first key recovery attack against WEP (known as the FMS attack) by exploiting the weaknesses of the Key Scheduling Algorithm used in RC4 [12]. They further identified certain classes of weak IVs which, if used, can enable the recovery of the private key from a relatively small number of observations. In 2004, the KoreK chopchop attack demonstrated the feasibility of breaking the WEP using the CRC32 checksum [13]. During the same year, KoreK introduced an enhanced set of attacks that break WEP without relying on weak IVs [14]. Later in 2005, Klein proposed a rather efficient approach for iteratively computing all of the secret key bytes, without the need of weak IVs as well [15]. The IEEE 802.11 fragmentation support was used in the Bittau attack, which appeared in 2006 [16]. In 2007, Pyshkin, Tews, and Weinmann (PTW) developed an enhanced version of the Klein attack by using key ranking techniques [17]. The PTW attack is considered to be the most efficient attack against WEP at the moment.

Careful inspection of passive WEP attacks reveals **their critical dependence** on collecting a large number of ciphertext/plaintext pairs with **unique IVs which are sent**

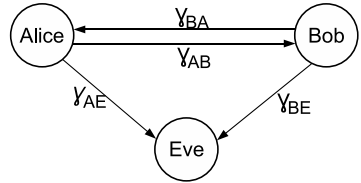

Fig. 1. System Model

**as plaintext**. For example, following aircrack-ng recommendation [19], an attacker would typically need 1.5 million frames with unique IVs, before launching a combined FMS and KoreK [14] passive WEP attacks[1]. Plaintext bytes could be guessed through the knowledge of the format of upper layer packets, e.g., ARP or IPv4 packets. Advanced statistical techniques can be used to recover the **only unknown variable**, i.e., the private key $K_s$, without much difficulty. In a nutshell, sending the IVs in the clear, along with using the same private $K_s$ in all frames, is the **main vulnerability** of the WEP protocol. Our proposed solution will transform the IVs into **secret keys** by exploiting the available ARQ mechanism in the 802.11 standard. This way, we overcome the main weakness of the WEP protocol while preserving the simplicity and user friendliness associated with using only one private key. Before proceeding to describing the detailed implementation of the proposed ARQ-WEP protocol, the next section develops the ARQ secrecy framework.

### III. KEYS THROUGH ARQ

Our main idea will be demonstrated in a simplified model with one transmitter (Alice), one legitimate receiver (Bob), and one passive eavesdropper (Eve). The underlying multi-path fading channels are modeled as binary time-varying block erasure channels. We assume a time-slotted system where time slots correspond to coherence intervals, such that each channel remains fixed over one time slot and changes randomly across time slots. Each ARQ epoch is transmitted in one time slot and if the frame is received, an ACK message is sent over the feedback link. A NACK message corresponds to no transmission (as stipulated by the 802.11 standard). Hence, an erasure event over the Bob-Alice channel will transform an ACK into a NACK. We let $\gamma_{XY}(i) \in [0, 1]$ denote the erasure probability at time slot $i$ for the channel connecting transmitter $X$ to receiver $Y$. The erasure events are assumed to be temporally and spatially independent while the erasure probabilities are assumed to be **random variables** which are correlated across space while being independent and identically distributed over time. The transmitter is assumed to know *a-priori* the joint distribution of the erasure probabilities but **not** their instantaneous values. The assumptions correspond to a Gaussian channel with independent additive noise, spatially correlated multi-path fading, no prior transmitter channel state

---
[1]An implementation for this set of attacks is available online in aircrack-ng toolset [19]

information (CSI), and perfect error detection. Figure 1 summarizes our modeling assumptions and the following result characterizes the maximum achievable **secure** key rate, under the assumption of noiseless feedback, i.e., $\gamma_{BA}(i) = 0 \ \forall i$,

*Theorem 1:* The secret key capacity for for the block erasure ARQ channel with noiseless feedback is

$$C_k = \mathbb{E}\left[(1 - \gamma_{AB})\gamma_{AE}\right] \quad (3)$$

*Proof:* The proof can be obtained as a special case of the main result in [7]. For the sake of completeness, however, the following is a sketch of the achievability argument. Alice wishes to share a secret key $W \in \mathcal{W} = \{1, 2, \cdots, M\}$ with Bob. Towards this end, Alice and Bob use an $(M, m)$ code consisting of : 1) a stochastic encoder $f_m(.)$ at Alice that maps the key $w$ to a codeword $x^m \in 2^m$, 2) a decoding function $\phi$: $2^m \to \mathcal{W}$ which is used by Bob to recover the key. Each codeword is partitioned into $k$ blocks each of $n_1$ symbols corresponding to one ARQ epoch ($m = k n_1$).

Alice starts by a random choice of the first block of $n_1$ symbols and sends it to Bob. If Bob successfully decodes this block, an ACK bit is sent to Alice. Otherwise, the frame is discarded by Bob and nothing is sent back. Upon the reception of an ACK bit, Alice stores the acknowledged frame, makes a random choice of the second block and sends it to Bob. If no ACK is received, Alice **replaces** the first block with another randomly chosen one and sends it to Bob. The process continues until Alice and Bob agree on $k$ blocks, each of $n_1$ symbols, corresponding to the private key. Using this approach, Eve does not benefit from its observation of the frames erased at Bob.

One can easily see that this protocol has transformed our ARQ channel into an erasure wiretap channel **without** feedback. In this equivalent model, the main channel is noiseless and the erasure probability at Eve is given by

$$\epsilon(i) = \gamma_{AE}(i)\left(1 - \gamma_{AB}(i)\right). \quad (4)$$

Under the assumption of an asymptotically large $k$, we apply the classical degraded wiretap channel results [1] to obtain (3). ∎

A few remarks are now in order
1) In the aforementioned security protocol, erasure events in the feedback channel will lead to mis-synchronization between Alice and Bob. This problem can be easily overcome at the expense of a larger overhead in the feedforward and feedback channel. Alice would include all the history of received ACK/NACK in each frame. Once an ACK is received, Alice will be assured that Bob has correctly received the past history. Alice will then flush the past history and will only include the recently received ACK/NACK messages in future transmissions. If we stop here, then the achievable key rate becomes

$$R_k = \mathbb{E}\left[(1 - \gamma_{AB})(1 - \gamma_{BA})\gamma_{AE}\right]. \quad (5)$$

The loss in rate, due to the dropped ACK frames, can be avoided if the feedback channel can accommodate more bits. In this case, Bob will resend all the previously dropped ACK bits along with the current one. Then, Alice will modify its past history accordingly and the same key will be generated by both nodes once synchronization is achieved. Clearly, this approach achieves the key rate reported in (3) when both $k$ and $n_1$ are asymptotically large, rendering the overhead loss negligible.

2) One may be tempted to assume that the noisy feedback from Bob to Eve will allow for increasing the secret key capacity. Unfortunately, Eve can easily overcome the loss of ACK bits via an exhaustive trial and error approach. More rigorously, since the ratio of feedback bits over feedforward bits is vanishingly small, the loss of ACK bits will not lead to an increase in the equivocation at Eve.

3) By invoking the classical wiretap channel results in our achievability argument, we implicitly assumed the use of a binning codebook [1]. Recent works have established the existence of Low Density Parity Check (LDPC) codes that achieve the secrecy capacity of the erasure wiretap channel under consideration in [8]–[10]. This approach, however, suffers from two drawbacks: 1) the complexity of LDPC coding/decoding can be prohibitive for many practical applications and 2) the delay, as measured by the number of successfully received ARQ epochs $k$, is asymptotically large. The following scheme avoids these limitations. The frames are chosen according to a i.i.d. uniform distribution over the binary alphabet $\{0, 1\}$ and the key is distilled as the modulo-2 sum of the $k$ successfully received frames at Bob. Clearly, this scheme enjoys a very low encoding/decoding complexity and, via the appropriate choice of $k$, one can limit the decoding delay. The choice of $k$ will, also, determine the operating point on the secrecy-throughput tradeoff curve. More specifically, Eve will be completely blind about the key if at least one of the $k$ frames is erased. Hence, the probability of **secrecy outage** is given by

$$P_{out} = \prod_{i=1}^{k}\left[1 - \gamma_{AE}(i)\right], \quad (6)$$

where $\gamma_{AE}(1), .., \gamma_{AE}(k)$ are i.i.d random variables drawn according to the distribution of Eve's channel. On the other hand, the average number of trials required to successfully transmit the $k$ ARQ frames to Bob is

$$\mathbb{E}[n] = k \, \mathbb{E}\left[\frac{1}{1 - \gamma_{AB}}\right], \quad (7)$$

resulting in the following key rate

$$R_k = \frac{1}{\mathbb{E}[n]} = \frac{1}{k \, \mathbb{E}\left[\frac{1}{1-\gamma_{AB}}\right]}. \quad (8)$$

Hence, choosing a larger $k$ will result in a smaller probability of secrecy outage at the expense of a lower key rate and a larger delay. If we account for erasure events in the feedback channel, with only ACK/NACK feedback, the achievable key rate becomes

$$R_k = \frac{1}{\mathbb{E}[n]} = \frac{1}{k\, \mathbb{E}\left[\frac{1}{(1-\gamma_{AB})(1-\gamma_{BA})}\right]}. \quad (9)$$

In the following section, the aforementioned modulo-2 key sharing scheme will be used to secure the WEP protocol.

## IV. THE ARQ-WEP PROTOCOL

Our proposed protocol conceals the Initialization Vectors (IVs) from Eve by introducing slight modifications to the currently implemented WEP protocol in 802.11 networks. Our goal is to prevent Eve from collecting the required number of IVs to launch her attack. This goal is achieved by seeding the RC4 algorithm with an IV that is *distributed* over all previously sent frames in a fashion that utilizes both the ARQ protocol and the independence between the channels seen by Eve and Bob. To formally describe the algorithm, we need the following notations. Let $Q(i) = 1$ if Alice receives an ACK for in $i^{th}$ frame ($Q(i) = 0$ otherwise) and $S(i) = 1$ if Bob successfully decrypts the $i^{th}$ frame ($S(i) = 0$ otherwise).

For data encryption, we apply the following modifications. We let the $i^{th}$ data frame carry a new randomly-generated IV in its MAC header, denoted by $V_h(i)$. However, for each data frame, the RC4 algorithm is seeded with the modulo-2 sum of all header IVs which were sent by Alice and successfully received by Bob. This sum is referred to as $V_e(i)$. As opposed to the original WEP protocol, i.e., (1) and (2), in the $i^{th}$ transmitted frame we have:

$$V_e(i) = \begin{cases} V_e(i-1) \bigoplus V_h(i-1), & \text{if } Q(i-1)=1\ ; \\ V_e(i-1), & \text{otherwise}. \end{cases}$$
$$(10)$$
$$C(i) = P(i) \bigoplus RC4\left(V_e(i), K_s\right), \quad (11)$$
$$A \to B : V_h(i), C(i), \quad (12)$$

where $V_e(0) = 0$. Bob attempts to decrypt the $i^{th}$ received frame with $K_s$ and the modulo-2 sum of all IVs perviously received, referred to as $V_d(i)$. If decryption fails, Bob excludes the last IV from the sum, i.e.,

$$V_d(i) = \begin{cases} V_d(i-1) \bigoplus V_h(i-1), & \text{if } S(i-1)=1\ ; \\ V_d(i-1), & \text{otherwise}. \end{cases}$$
$$(13)$$

Again, $V_d(0) = 0$. Furthermore, the history of all received ACKs by Alice is embedded in each encrypted frame. This way, we avoid any mis-synchronization that could happen due to the loss of an ACK frame; without any additional feedback bits.

Now, in order to launch an attack, Eve attempts to collect as many of the data frames sent by Alice as possible. In our scheme, however, the usefulness of the collected traffic depends on Eve's ability to correctly compute $V_e$ for each received frame. Such ability is hampered as Eve becomes completely **blind** upon missing a **single** ACKed frame. This observation motivates the use of a number of *initialization* frames at the beginning of each session (before any data exchange), to reduce the secrecy outage probability by adding more IVs to the encryption sum. The initialization frames contain only IVs, as plaintext, so as to reveal no information about the secret key, $K_s$. The experimental results, reported in the following section, will investigate the throughput-secrecy tradeoff governed by the ratio of the total size of initialization frames to the session size, in different practical settings. **In summary, the IV used for encryption/decryption is the secret key shared via the underlying ARQ protocol.**

Finally, we report some of the implementation details. Our ARQ-WEP prototype was incorporated in the madwifi-ng driver [20] by modifying the wlan_wep and ath_pci modules, in software encryption mode. The detection of acknowledgments and timeout events was established by using the Hardware Abstraction Layer (HAL) reports to the driver. In an infra-structure network architecture, the Access Point (AP) and each client store all the necessary information for data exchange. The eavesdropper has to maintain similar information for each client/AP session of interest. Initialization frames are implemented as (un-encrypted) association management frames with extended subtypes. To optimize performance, these frames are exchanged in bursts with the use of custom NACKs. The average initialization frame length is 42 bytes, which is negligible, as compared to a typical data frame size. The total number of initialization frames varies depending on the required secrecy level and acceptable overhead.

## V. EXPERIMENTAL RESULTS

We deployed the modified madwifi-ng driver on laptops running the FC8 Linux distribution and D-Link wireless cards (DWL-G650). We ran the experiments in an infra-structure IEEE 802.11g network composed of an AP and a single client (STA), with one passive eavesdropper, enabled in monitor mode.

We first evaluate the expected number of *useful* frames that Eve obtains per session, i.e., the *data* frames that Eve could successfully compute their encryption IVs. For each session between Alice and Bob, the expected number of these frames can be *upper bounded* as

$$\mathbb{E}[u] \leq \frac{\mathbb{E}[\acute{\gamma}_{AE}]^{k_i+1} - \mathbb{E}[\acute{\gamma}_{AE}]^{k+1}}{\mathbb{E}[\gamma_{AE}]}, \quad (14)$$

where $\acute{\gamma}_{AE} = 1 - \gamma_{AE}$, $k_i$ is the number of initialization frames successfully received by Bob, and $k$ is the total number of frames successfully received by Bob. As shown in (14), a slight increase of the overhead introduced by the initialization frames, results in a significant decrease in the number of frames Eve could use per session, and thus, a significant increase in the listening time needed to launch an attack.

We validated our analytical estimate **experimentally** by generating one-way traffic between the AP (Alice) and the client node (Bob). We equipped Eve's driver with the same logic used in our protocol, i.e. the modified driver monitors all transmitted frames in the network, extracts their IVs, and sums them based on the observed ACKs/timeouts. We launched two experiments in two different environments. In the first, Eve was observed to have better channel conditions, on the average, than Bob. While in the second, the situation was reversed and all channels suffered from relatively large erasure probabilities. We ran each experiment at different numbers of initialization frames, compared the IVs obtained by Eve and Bob, and calculated the average number of useful frames at Eve, over 40 trials for each sample point. For both experiments, the data session size is 100000 frames.

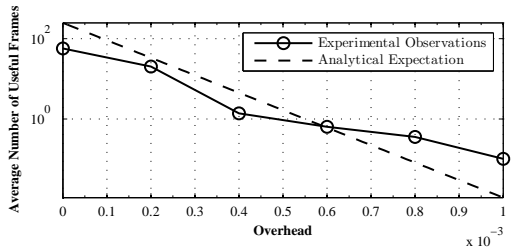

Fig. 2. The average number of useful frames obtained by Eve. $\mathbb{E}[\gamma_{AB}] = 0.005$, $\mathbb{E}[\gamma_{BA}] = 0.009$ and $\mathbb{E}[\gamma_{AE}] = 0.004$

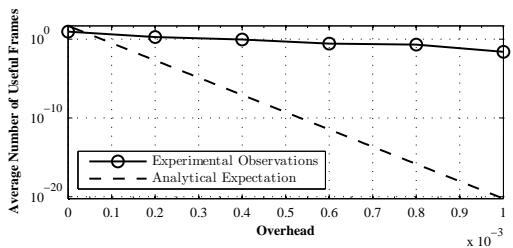

Fig. 3. The average number of useful frames obtained by Eve per session. $\mathbb{E}[\gamma_{AB}] = 0.01$, $\mathbb{E}[\gamma_{BA}] = 0.01$ and $\mathbb{E}[\gamma_{AE}] = 0.02$

The results are reported in Figures 2 and 3. The disagreement between the analytical estimates and the experimental results appears to be due to the small number of samples used in our experiments. However, to compare our secrecy gain relative to the original WEP, one can use the reported results in these figures to estimate the required time for Eve to gather the required 1.5 million frames to launch an attack. Under the standard WEP operation, we optimistically assume that Eve needs 10 minutes to gather such traffic. On the other hand, the estimated average attack time, with the proposed ARQ-WEP and no initialization overhead, is 19.35 hours and 5.23 days, for the first and second setups, respectively. An overhead of 0.001 extends the required average listening time to 1.24 years and 5.07 years, respectively. **Clearly, the ARQ-WEP is able to achieve very impressive secrecy gains with only a marginal loss in throughput**.

## VI. CONCLUSION

To the best of our knowledge, this paper constitutes the first attempt at securing practical wireless networks using information theoretic principles. The proposed ARQ-WEP protocol was able to avoid the main vulnerability of the WEP protocol, namely, the transmission of the IV header as a clear text, via only minor modifications of the existing WEP encryption mechanism. The underlying idea is to benefit from ARQ feedback and multi-path fading to create an advantage for the legitimate receiver over potential eavesdroppers. Our experimental results demonstrate the significant security advantage offered by the ARQ-WEP protocol over the traditional implementation of the WEP protocol; e.g., a passive attack that requires 10 minutes with the traditional WEP protocol will need around a year of listening time with the ARQ-WEP protocol under realistic implementation assumptions. The most interesting aspect of our work is, perhaps, the development of an information theoretic security protocol that *enhances* the security of an existing security protocol, i.e., the WEP, at the expense of a minimal additional complexity.